\documentclass[12pt,preprint]{aastex}
\usepackage[]{natbib}
\usepackage{graphicx, amsmath}
\usepackage{color}

\begin{document}

\title{IDCS J1433.2+3306: An IR-Selected Galaxy Cluster at z = 1.89}
\author{Gregory R. Zeimann\altaffilmark{1}, S. A. Stanford\altaffilmark{1,2}, Mark Brodwin\altaffilmark{3}, Anthony H. Gonzalez\altaffilmark{4}, Gregory F. Snyder\altaffilmark{5}, Daniel Stern\altaffilmark{6}, Peter Eisenhardt\altaffilmark{6}, Conor Mancone\altaffilmark{4}, and Arjun Dey\altaffilmark{7}}
\altaffiltext{1}{Department of Physics, University of California, One Shields Avenue, Davis, CA 95616}
\altaffiltext{2}{Institute of Geophysics and Planetary Physics, Lawrence Livermore National Laboratory, Livermore, CA 94550}
\altaffiltext{3}{Department of Physics, University of Missouri, 5110 Rockhill Road, Kansas City, MO 64110}
\altaffiltext{4}{Department of Astronomy, University of Florida, Gainesville, FL 32611}
\altaffiltext{5}{Harvard-Smithsonian Center for Astrophysics, 60 Garden Street, Cambridge, MA 02138}
\altaffiltext{6}{Jet Propulsion Laboratory, California Institute of Technology, Pasadena, CA 91109}
\altaffiltext{7}{NOAO, 950 North Cherry Avenue, Tucson, AZ 85719}


\begin{abstract}
We report the discovery of an IR-selected galaxy cluster in the IRAC Distant Cluster Survey (IDCS). New data from the {\it Hubble Space Telescope} spectroscopically confirm IDCS J1433.2+3306 at $z = 1.89$ with robust spectroscopic redshifts for seven members, two of which are based on the 4000\AA\ break.  Detected emission lines such as [O\textsc{ii}] and H$\beta$ indicate star formation rates of $\gtrsim 20$ M$_{\odot}$ yr$^{-1}$ for three galaxies within a 500 kpc projected radius of the cluster center.   The cluster exhibits a red sequence with a scatter and color indicative of a formation redshift $z$$_f$ $\gtrsim$ 3.5.  The stellar age of the early-type galaxy population is approximately consistent with those of clusters at lower redshift (1 $< z <$ 1.5) suggesting that clusters at these redshifts are experiencing ongoing or increasing star formation.   
\end{abstract}

\section{Introduction}

Emerging from the cosmic web, galaxy clusters are the most massive, bound configurations in the universe and provide insights into the formation and growth of large-scale structure as well as the physics that drives galaxy evolution.  The past several decades have seen an increase in the size and fidelity of cluster samples, which have enabled measurements of their abundance and evolution, offering important constraints on cosmology (see \citealp{allen2011}).  Moreover, galaxy clusters even at $z$ $\gtrsim$ 1 harbor a high density of old, massive stellar populations (\citealp{eisenhardt2008,snyder2012}), serving as excellent laboratories for studying the formation and evolution of stars in the earliest collapsed structures.  Galaxy clusters at $z$ $>$ 1.5  provide an early glimpse of the stellar mass build-up in rich, highly biased environments (e.g., \citealp{mancone2010}), and, if massive enough, may even have cosmological implications (\citealp{mortonson2011}).      
 
Pushing cluster surveys beyond $z = 1.5$ has been a challenge for typical methods such as X-ray, Sunyaev-Zel'dovich (SZ), and color-selection techniques.  Mass sensitivity issues plague both X-ray and SZ cluster survey methods as they cannot yet reach the required mass limits over sizable areas to discover large samples of $z > 1.5$ clusters.  Red-sequence selection, one of the more popular color-selection methods, begins to fail in the optical at $z\sim1.2$ as the 4000\AA\ break exits the wavelength coverage of CCD imagers and spectrographs.  A logical extension of this method to the infrared (IR) using the {\it Spitzer} IRAC imager has been successful at $z < 1.5$ (\citealp{muzzin2009}), but may be disadvantaged at $z > 1.5$ as the specific star formation densities in clusters become comparable to the field and density contrasts become harder to identify (\citealp{romeo2005,tran2010,brodwin2012}).  \citet{papovich2008} used an IR color to effectively select high-redshift galaxies which was later used to identify a galaxy cluster at $z = 1.62$ (\citealp{papovich2010})\footnote[8]{This cluster was also independently discovered in the X-ray by \citet{tanaka2010}.}.  Despite the challenges, other spec-z confirmed clusters at $z > 1.5$ have been discovered (\citealp{fassbender2011, santos2011, gobat2011}).  

The stellar-mass selected IRAC Shallow Cluster Survey (ISCS; \citealp{eisenhardt2008}) provides a variation on the traditional techniques.  Identifying cluster candidates as 3-D spatial overdensities (RA, Dec, and photometric redshift), the ISCS compiled a catalog of over 300 candidates in $\sim$7.25 deg$^2$ spanning 0 $<$ $z$ $<$ 2, including more than 100 at $z$ $>$ 1.  More than 20 clusters at 1 $<$ $z$ $<$ 1.5 have been spectrosopically confirmed to date (\citealp{stanford2005, brodwin2006, elston2006, eisenhardt2008, brodwin2011, brodwin2012}).    

With the emphasis of pushing to $z$ $>$ 1.5, the IRAC Distant Cluster Survey (IDCS) is an extension of the ISCS, drawing from the NOAO Deep Wide-Field Survey (NDWFS; \citealp{jannuzi1999}), the {\it Spitzer} Deep, Wide-Field Survey (SDWFS; \citealp{ashby2009}), and NIR data from the NOAO Extremely Wide-Field Infrared Imager (NEWFIRM).   IDCS selects galaxy cluster candidates as 3-D spatial overdensities using a 4.5$\mu$m selected sample with robust photometric redshifts ($B$$_w$$R$$I$[3.6][4.5]; \citealp{brodwin2006}).  One such candidate has already been spectroscopically confirmed at $z$$=$1.75 and is the most massive cluster known at $z > 1.4$ (\citealp{stanford2012, brodwin2012a, gonzalez2012}).  Here we report the spectroscopic confirmation of another candidate which had a photometric redshift estimate of $z$ $=$ 1.8: IDCS J1433.2+3306 is at $z$$=$1.89. 

We present the optical and near-infrared imaging data in \S2, the spectroscopic observations and resulting redshifts in \S3 and \S4, respectively, and the analysis of the galaxy population in \S5. We use Vega magnitudes and a WMAP7+BAO+$H_0$ $\Lambda$CDM cosmology (\citealp{komatsu2011}): ${\Omega}_{M}$ = 0.272, ${\Omega}_{\Lambda}$ = 0.728, and $H_0$ = 70.4 km s$^{-1}$ Mpc$^{-1}$.

\label{sec:intro}

\section{Optical and Near-IR Imaging}
\label{sec:2}

Following the procedure described in \citet{eisenhardt2008}, IDCS J1433.2+3306 was selected as a 3-D overdensity (RA, Dec, $z_{\rm phot}$) with a photometric redshift of $z$=1.8 (see Table 2 for the list of photometric candidate members).  Photometric redshifts were determined in a similar fashion to \citet{brodwin2006}, using \citet{polletta2007} empirical templates fit to 4$\arcsec$ aperture fluxes from a combination of optical (NDWFS) and infrared (SDWFS) data.    A color image ($B_w$, $I$, and $[4.5]$) shown in Figure 1 shows a clear overdensity of red galaxies with a filamentary-like structure.  At $\sim$3.5 Gyr after the big bang, many galaxy clusters are likely to be found in the formation process with looser, less bound structures.      

Using the {\it Hubble Space Telescope} ({\it HST}), deep optical and NIR follow-up imaging was obtained with ACS and WFC3.  Optical imaging in F814W with ACS was acquired in a single pointing consisting of 8 x 564 s exposures.  NIR imaging in F160W with WFC3 was obtained in two slightly overlapping pointings, each comprised of 2 x 350 s dithered exposures.  Both the ACS and WFC3 data were reduced using standard procedures with the MultiDrizzle software (\citealp{fruchter2009}).  A pseudo-color image using F814W and F160W is shown in the left panel of Figure 2 along with zoomed in cutouts of spectroscopic members and red sequence candidates in the right panel.

All of the {\it HST} imaging was registered to the same pixel scaling (0.065$\arcsec$ per pixel) using \textsc{Scamp/SWarp} (\citealp{bertin2002,bertin2006}).  Source extraction was performed in dual image mode using \textsc{SExtractor} (\citealp{bertin1996}) with the WFC3 filter as the detection band.  Colors were measured in 0.8$\arcsec$ diameter apertures while ``total'' magnitudes were measured using MAG AUTO.  The photometric uncertainty (which is dominated by sky shot noise, ${\sigma}_{\rm sky}$) was estimated by randomly placing 5000 0.8$\arcsec$ diameter apertures in the image and measuring the width of the negative half of the roughly Gaussian distribution to avoid contamination from real sources.  For verification, the correct scaling of the photometric error was found when this method was applied to the individual dithered images and then to the final co-add.   

Morphologies were classified using a S\'{e}rsi\c{c} profile.  Running in succession, \textsc{Galapagos} and \textsc{GALFIT} (\citealp{haussler2011,peng2010}) determined the morphologies of galaxies in the WFC3 image.  \textsc{Galapagos} provided an initial guess for \textsc{GALFIT} which then fit a single S\'{e}rsi\c{c} profile to every galaxy.  The S\'{e}rsi\c{c} index, $n$, is used to classify galaxy morphologies as either being early-type ($n$ $>$ 2.5) or late-type ($n$ $\le$ 2.5).

\section{Spectroscopy}
\label{sec:3}

\subsection{Keck/LRIS Optical Spectra}

Deep, optical spectroscopy was obtained for IDCS J1433.2+3306 using the Low-Resolution Imaging Spectrograph (LRIS; \citealp{oke1995}) on the 10-m Keck I Telescope during the nights of UT 2010 May 11-12.  We used slit masks with 1.1'' x 10'' slitlets, the 400/8500 grating on the red side, the D680 dichroic, and the 400/3400 grism on the blue side.  Night sky conditions were clear with 0.7$\arcsec$ seeing.   Exposures of 3 x 1800s were taken on both nights.  Each slitlet was reduced individually using standard reduction techniques, which included correction for the relative spectral response calibrated using longslit observations of Wolf 1346 and Feige 34 (\citealp{massey1990}) observed on the same nights at the parallactic angle.\\

\subsection{{\it HST}/WFC3 Grism Spectra}

We used {\it HST}/WFC3 to take grism spectroscopy of IDCS J1433.2+3306.  We used the two IR grisms: G102 and G141. These grisms have a throughput greater than 10$\%$ in the range of 0.80 - 1.15$\mu$m and 1.08 - 1.69$\mu$m, respectively, providing a continuous wavelength coverage of nearly 1$\mu$m, ideal for the identification of common galaxy spectral features over a wide range of redshifts.  The spectral resolution for the G102 and G141 grisms are 49 \AA\ and 93 \AA, respectively, which is sufficient to securely identify cluster members with a typical redshift accuracy of ${\sigma}_z$ $\approx$ 0.01.   

For slitless spectroscopy, a direct image is a necessary companion to the grism image in order to calibrate the wavelength scale and properly extract the spectra.  We chose broadband filters that closely matched the grism spectral coverage: F105W for G102 and F140W for G141 (see Figure 3).  The sizes and positions of detected objects in the direct image were used to establish the location, wavelength scale, and extraction windows for spectra in the grism images.  Reductions were performed using the aXe software (\citealp{kummel2009}).  Using the best available calibration files\footnote[9]{http://www.stsci.edu/hst/wfc3/analysis/grism\_obs/calibrations/},  spectra were extracted following the steps found in the WFC3 Grism Cookbook\footnote[10]{http://www.stsci.edu/hst/wfc3/analysis/grism\_obscookbook.html}.  A more detailed description of the grism data analysis for IDCS J1433.2+3306 and seventeen additional $z > 1$ clusters will be presented in an upcoming paper.

\section{Redshift Measurements}
\label{sec:4}

Although the Keck spectra are fairly deep, no secure redshifts were identified.  Many of the spectra were either featureless noisy continua or were absent of any emission.

For ease of visual inspection, the reduced WFC3 grism spectra were displayed in a webpage format using aXe2web.\footnote[11]{http://axe.stsci.edu/axe/axe2web.html\#ref\_1}  This allowed examination of both the individual 2-d grism spectrum as well as the 1-d extraction.  Emission lines were identified by eye and inspected in detail.  A redshift quality scale was used to quantify the robustness of the measurement.  A spectrum exhibiting a single feature was given a quality value of Q$=$C, while a spectrum showing two features consistent with the same redshift was assigned a quality value of Q$=$B, and a spectrum with three or more features indicating a single redshift was denoted with a quality value of Q$=$A.  Robust redshifts, Q$=$A or Q$=$B, from emission line detections were found for five cluster members and their spectra are shown in Figure 4.  

Additionally, galaxies exhibiting red continua and a distinctive break were also examined in detail.  These early-type galaxy candidates were cross-matched with a sample of templates spanning a range of ages, metallicities, and extinctions.  We used EzGal\footnote[12]{http://www.baryons.org/ezgal/} and \citet[2007 version]{bruzual2003} to produce composite stellar population models with a Chabrier initial mass function (IMF; \citealp{chabrier2003}) and an exponentially-declining star formation rate ($\tau = 0.1$ Gyr) modified using the \citet{calzetti2000} extinction law.  Searching a grid (value $=$[start:step:end]; age $=$[0.1:0.1:3.0]Gyr, $A_V=$[0.0:0.1:2.0], $Z=$[0.4,1.0,2.5]$Z_{\odot}$, and $z=$[1.50:0.01:2.50]) for each early-type galaxy candidate, a best-fit template and redshift was found using a ${\chi}^{2}$-minimization method.  A redshift was considered to be robust if the best-fit template redshift was consistent with the identifiable D4000 break.  Spectra of the two cluster members identified with robust redshifts using this method are plotted in Figure 5.  Red sequence candidates (see Figures ~\ref{fig:f2}, ~\ref{fig:f7}, ~\ref{fig:f8}) without robust redshifts were either too faint to identify a D4000 break, strongly affected by contamination from overlapping spectra, or had a best-fit model redshift that did not match the identified D4000 break.

The redshift histogram, plotted in Figure 6, shows the results from both the emission line identification and the template cross-matching.  For the purpose of this paper, we adopt the cluster definition of \citet{eisenhardt2008} which holds that a cluster is confirmed if there are at least five cluster members within a radius of 2 Mpc whose spectroscopic redshifts match to within $\pm2000$($1+ \langle z_{\rm spec}\rangle$) km s$^{-1}$.  In IDCS J1433.2+3306, there are seven galaxies with robust redshifts at $z$$=$1.89$\pm$0.01 all within a radius of 1 Mpc, thus spectroscopically confirming the galaxy cluster at $z$ = 1.89.

\section{Galaxy Population}
\label{sec:5}

\subsection{Early-Type Galaxies}

 We used a combination of colors, magnitudes, and morphologies to investigate the early-type galaxy (ETG) population.  Figure 7 presents a color-magnitude diagram (CMD) using the deep optical and NIR imaging of ACS and WFC3 (FoV of 7.7 arcmin$^2$).  To quantify the red sequence, we employed the multi-step procedure from Snyder et al. (2012) which starts with a single burst formation model of the Coma cluster color-magnitude relation (CMR) derived in \citet{eisenhardt2007} and a formation epoch of $z$$_f$$=$3.  The colors of this model at $z = 1.89$ were then subtracted from every galaxy's color, which resulted in a quantity denoted as $\Delta$, as shown in Figure 8.  Red sequence candidate galaxies were then selected from the color region -0.5$<$$\Delta$$<$1.0, and galaxies fainter than $H^{\star}$$+$1.5 were removed (where $H^{\star}$  is evolved from Coma ETGs).  Most of the galaxies in the selected magnitude and color region lack spectroscopic or photometric redshifts, so a morphological criterion was applied, $n$ $>$ 2.5, to reduce the number of interlopers and select ETGs.  Finally, we rejected sources with colors more than two median absolute deviations from ${\Delta}_{0}$, the color zeropoint as discussed below.  Sources discarded in the last step are shown as gray dots in Figure 8 while the remaining objects in the red sequence selection are represented by filled red circles.  Identifications and positions for all filled red circles in Figure 8 are shown in Table 3 along with their photometric redshifts when available. 
 
To measure the color zeropoint, intrinsic scatter, and associated uncertainties of this red sequence we used the biweight estimates of location and scale (\citealp{beers1990}). We calculated the intrinsic scatter ${\sigma}_{\rm int}$ in the red sequence sample by subtracting in quadrature the median color error from the biweight scale estimate.  To estimate associated uncertainties we performed these calculations on 1000 bootstrap resamplings and measured the median absolute deviations of the color zeropoint ${\Delta}_{0}$ and intrinsic scatter ${\sigma}_{\rm int}$ distributions.  Plotted in Figure 8 are red horizontal lines offset a distance ${\sigma}_{\rm int}$ above and below the color zeropoint ${\Delta}_{0}$ of the red sequence.  From this method, the morphologically-selected red sequence sample has a color zeropoint ${\Delta}_{0}$ of F814W$-$F160W$=$4.09$\pm$0.07 and intrinsic scatter of 0.21$\pm$0.05 mag; transformed to the rest frame ($U-V$)$_0$ the intrinsic scatter is 0.10$\pm$0.02.  The color is consistent with a single-burst formation epoch of 4 $\lesssim$ $z_{f}$ $\lesssim$ 6, while the scatter is more consistent with 3.5 $\lesssim$ $z_{f}$ $\lesssim$ 4.5, derived from a simple passive evolution model (\citealp[2007 version]{bruzual2003}).  When restricted to the same parameter values of the red sequence analysis (i.e., solar metallicity, zero extinction, the best-fit redshift, and a Chabrier IMF) \citet[2007 version]{bruzual2003} best-fit models of the two ETG grism spectra indicate a formation epoch of 4 $\lesssim$ $z_{f}$ $\lesssim$ 7 and 3 $\lesssim$ $z_{f}$ $\lesssim$ 4 for J143316.5+330716 and J143311.5+330639, respectively.\footnote[13]{We fit models to 1000 realizations of the grism spectra assuming Gaussian random errors and used the 16$^{th}$ and 84$^{th}$ percentile values of the best-fit age to represent the range of formation redshifts.}  While formed at a higher redshift, the stellar age of the ETG population in IDCS J1433.2+3306 is roughly consistent with those inferred in the lower redshift clusters of the ISCS at 1 $< z <$ 1.5 (\citealp{snyder2012}).

\subsection{Star Formation and Active Galaxies}

Near the peak of galaxy formation activity, a cluster at $z$ $=$ 1.89 is expected to have signs of both high rates of star formation ($>$10 M$_{\odot}$ yr$^{-1}$) and active galactic nuclei (AGN).  We detected five emission line galaxies, each exhibiting some combination of [O\textsc{ii}]$\lambda$3727, H$\beta$, and [O\textsc{iii}]$\lambda$$\lambda$4959,5007.  These emission lines are likely due to high rates of star formation, AGN activity, or a combination of the two.  Since typical diagnostic emission lines (e.g., H$\alpha$ and [N\textsc{ii}]) fall outside of the WFC3 grism wavelength coverage at this high redshift, we used X-ray detection in XBo\"{o}tes (\citealp{murray2005,kenter2005}) and the empirical mid-IR criteria from \citet{stern2005} to distinguish AGN from star-forming galaxies.  One of the five emission-line cluster members, J143313.1+330736, was detected in the X-ray, exhibiting a flux F$_{X}$ $=$ (4.35$\pm$0.89) $\times$ 10$^{-15}$ ergs s$^{-1}$ cm$^{-2}$ in the 0.5$-$7.0 keV range and a hardness ratio of HR $= -0.20\pm0.14$ (HR defined in \citealp{kenter2005}).  Assuming an intrinsic photon index of $\Gamma = 1.9$, the HR suggests a hydrogen column density in the range of $N_H = 10^{22} - 10^{23}$ cm$^{-2}$ (\citealp{wang2004}).  As well as being detected in the X-ray data, the IRAC colors of J143313.1+330736 indicate that it is an AGN and the morphology suggests a merging system (\#7 in the right panel of Figure ~\ref{fig:f2}).  The existence of an AGN in IDCS J1433.2+3306 is not surprising given the increased incidence of AGN seen in $z > 1$ clusters (\citealp{galametz2009}).  The other four emission-line cluster members were not detected in the X-ray and were too faint to test the mid-IR selection technique.  In subsequent discussion, we will assume that all of the emission in the remaining four  galaxies is due to star formation activity.

The most reliable star formation indicator is H$\alpha$, but this unfortunately falls outside of our wavelength coverage. Instead, we use H$\beta$ and [O\textsc{ii}], when available, to calculate star formation rates.  To convert H$\beta$ luminosity into a star formation rate we use an assumed ratio (H$\alpha$/H$\beta$) = 2.86 from case B recombination (\citealp{storey1995}) and the H$\alpha$ star formation law of \citet{kennicutt1998}.   For the case of [O\textsc{ii}] emission, we use the star formation law from equation 15 of \citet{kewley2002}.  For simplicity, we assume an extinction of $A_V$ = 1.0 using \citet{calzetti2000}, and account for stellar absorption via a 20$\%$ additive correction for H$\beta$.  To measure the line strength we fit a third order polynomial to the continuum, excluding regions of emission.  We then fit a Gaussian to the continuum-subtracted emission line, correct the measured flux for extinction and stellar absorption, and convert it into a luminosity and star formation rate.  The errors from this process are dominated by our assumptions with an estimated uncertainty of a factor of 2.  The derived star formation rates are in Table 1.  Three of the four galaxies have high star formation rates ($\gtrsim$ 20 M$_{\odot}$ yr$^{-1}$) and lie within 500 kpc of the cluster center.      
 
\section{Summary and Discussion}
\label{sec:6}

IDCS J1433.2+3306 was selected as a 3-D overdensity (RA, Dec, $z_{\rm phot}$) with a photometric redshift of $z$=1.8.  Using the WFC3 grism, we have spectroscopically confirmed seven cluster members, $\langle z_{\rm spec}\rangle$$=$1.89, within a projected radius of 500 kpc.  The confirmed members include both emission-line and early-type galaxies.  One of the five emission-line galaxies shows signatures of AGN activity including X-ray emission.  Using detected [O\textsc{ii}] and H$\beta$ emission, three of the remaining four emitters have star formation rates consistent with $\gtrsim$20 M$_{\odot}$ yr$^{-1}$.           

To understand the early-type galaxy population, we performed a red sequence selection with moderately deep ACS and WFC3 imaging and the inclusion of morphological information (S\'{e}rsi\c{c} index).  The color and intrinsic scatter of the measured red sequence indicated a single-burst formation epoch of 4 $\lesssim$ $z_{f}$ $\lesssim$ 6 and 3.5 $\lesssim$ $z_{f}$ $\lesssim$ 4.5, respectively.  When restricted to the same parameter values of the red sequence analysis (i.e., solar metallicity, zero extinction, the best-fit redshift, and a Chabrier IMF) \citet[2007 version]{bruzual2003} best-fit models of the two ETG grism spectra indicate a formation epoch of 4 $\lesssim$ $z_{f}$ $\lesssim$ 7 and 3 $\lesssim$ $z_{f}$ $\lesssim$ 4 for J143316.5+330716 and J143311.5+330639, respectively.  Both the red sequence analysis and the grism spectra indicate a  similar formation epoch and, while formed at a higher redshift, the stellar age of the ETG population in IDCS J1433.2+3306 is roughly consistent with those inferred in clusters of the ISCS at 1 $< z <$ 1.5 (\citealp{snyder2012}).  The approximately constant stellar age of the ISCS/IDCS cluster sample from 1 $< z <$ 1.89 supports the conclusions in \citet{snyder2012} that clusters at these redshifts are experiencing ongoing or increasing star formation.  

While IDCS J1433.2+3306 satisfies the criteria for a cluster given in \citet{eisenhardt2008}, it does not show a centralized concentration of galaxies like that of lower redshift clusters, and an assessment of its dynamical state will require further follow-up.  The highest priority observations would measure the velocity structure of the cluster.

If IDCS J1433.2+3306 is representative of the average cluster in the ISCS sample, an extrapolation of the clustering analysis in \citet{brodwin2007} to $z > 1.5$ suggests the mass of IDCS J1433.2+3306 is M $\sim 10^{14} M_{\odot}$.   Follow-up at X-ray energies of sufficient depth ($\gtrsim$100 ks), may allow for the detection of the intra-cluster medium (ICM) and assessment of the maturity of the cluster.  Surface brightness dimming makes detection of the ICM difficult in the X-ray at very high redshifts; however, a galaxy cluster at a similar mass and redshift ($z = 2.09$; \citealp{gobat2011}) has been detected as an extended X-ray source with {\it Chandra} and {\it XMM} in a total exposure time of $\sim$160ks and has yielded a mass measurement (M $= (5-8)\times10^{13} M_{\odot}$).  An alternative detection of the ICM would be through the SZ technique as was done for IDCS J1426.5+3508 (\citealp{brodwin2012a}), but the feasibility of this method strongly depends on the mass of IDCS J1433.2+3306.  Identifying more members with ground-based NIR spectroscopy may allow for a velocity dispersion measurement.  With the newly commissioned Multi-Object Spectrometer for Infra-Red Exploration (MOSFIRE on Keck), confirming new cluster members is an exciting possibility.\\

AHG acknowledges support from the National Science Foundation through grant AST-0708490. This work is based in part on observations made with
the {\it Spitzer Space Telescope}, which is operated by the Jet Propulsion Laboratory, California Institute of Technology under a contract with NASA. Support for this work was provided by NASA through an award issued by JPL/Caltech. Support for {\it HST} programs 11663 and 12203 were provided by NASA through a grant from the Space Telescope Science Institute, which is operated by the Association of Universities for Research in Astronomy, Inc., under NASA contract NAS 5-26555. Some of the data presented herein were obtained at the W. M. Keck Observatory, which is operated as a scientific partnership among the California Institute of Technology, the University of California and the National Aeronautics and Space Administration. The Observatory was made possible by the generous financial support of the W. M. Keck Foundation. This work makes use of image data from the NOAO Deep WideÐField Survey (NDWFS) as distributed by the NOAO Science Archive. NOAO is operated by the Association of Universities for Research in Astronomy (AURA), Inc., under a cooperative agreement with the National Science Foundation. 

We thank the anonymous referee for their constructive comments, Matt Ashby for creating the IRAC catalogs for SDWFS, Buell Jannuzi for his work on the NDWFS, Michael Brown for combining the NDWFS with SDWFS catalogs, and Steve Murray and the XBo\"{o}tes team for obtaining the {\it Chandra} data in the Bo\"{o}tes field. This paper would not have been possible without the efforts of the support staffs of the Keck Observatory, {\it Spitzer Space Telescope}, {\it Hubble Space Telescope}, and {\it Chandra X-ray Observatory}.  The work by SAS at LLNL was performed under the auspices of the U. S. Department of Energy under Contract No. W- 7405-ENG-48.

\bibliography{biden}

\begin{figure*}[htp]
\includegraphics[totalheight=.48\textwidth,width=.48\textwidth]{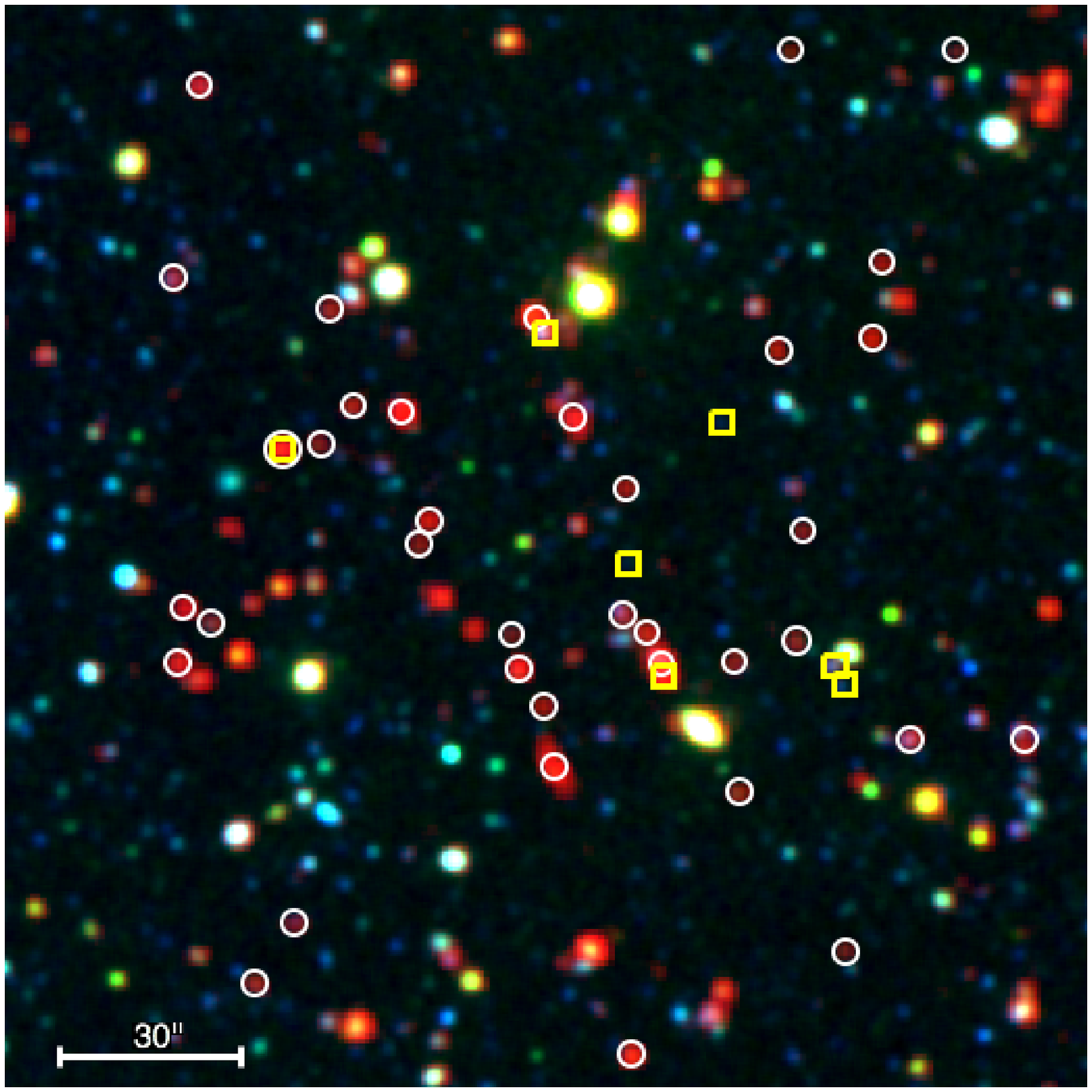}
\centering
\caption{ Pseudo-color image of IDCS J1433.2+3306, 3' on a side, constructed using NDWFS B$_{W}$ and I, and SDWFS [4.5].  North is up and east is to the left.  The yellow boxes are spectroscopically confirmed members, the white circles are candidate members based on their photometric redshift ($1.6 < z_{\rm phot} < 2.0$), and a 30'' ($\sim$260 kpc) scale bar is given for reference.}\label{fig:f1}
\end{figure*}

\begin{figure*}[htp]
\fbox{\includegraphics[totalheight=.48\textwidth,width=.48\textwidth]{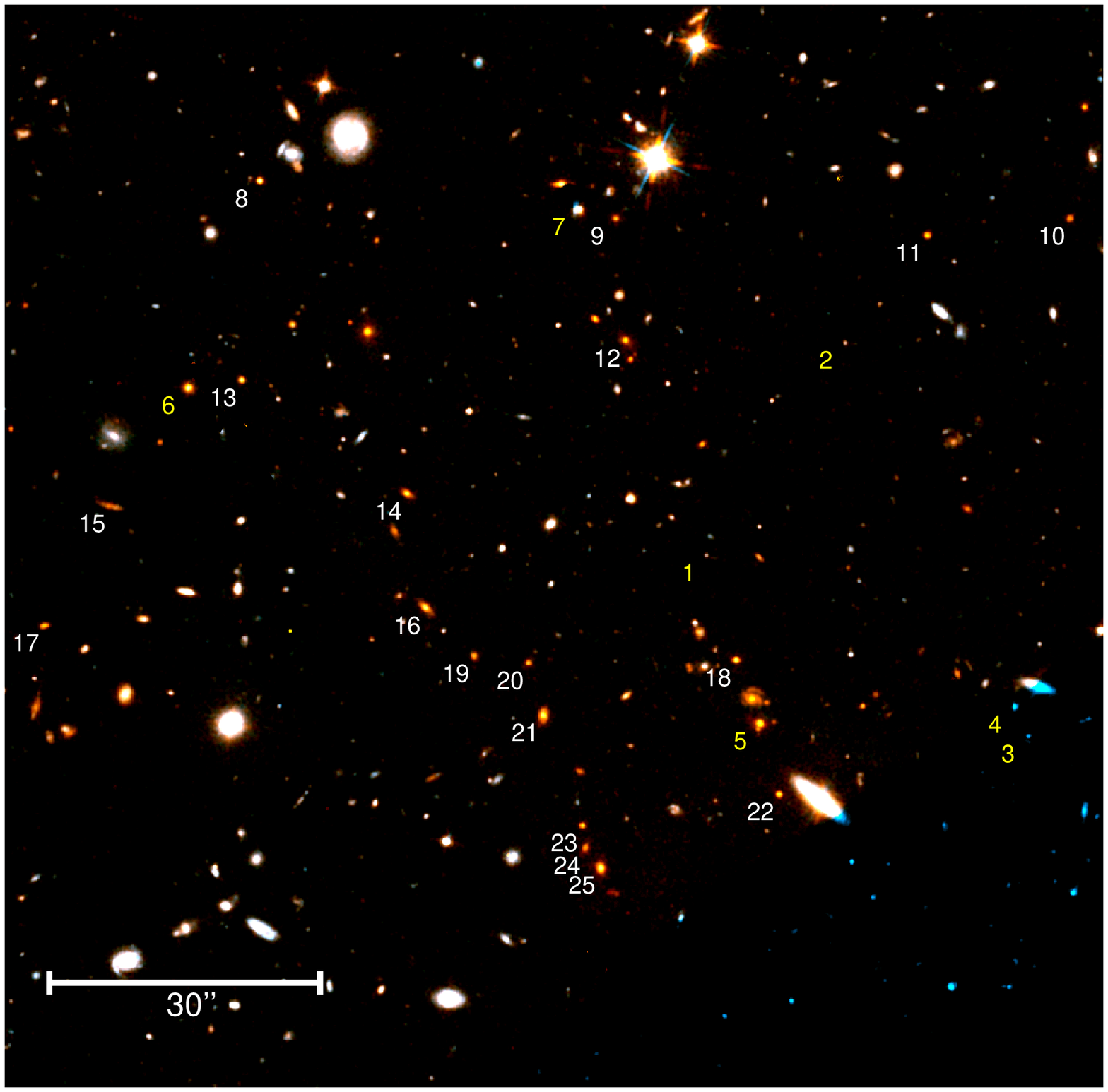}}
\fbox{\includegraphics[totalheight=.48\textwidth,width=.48\textwidth]{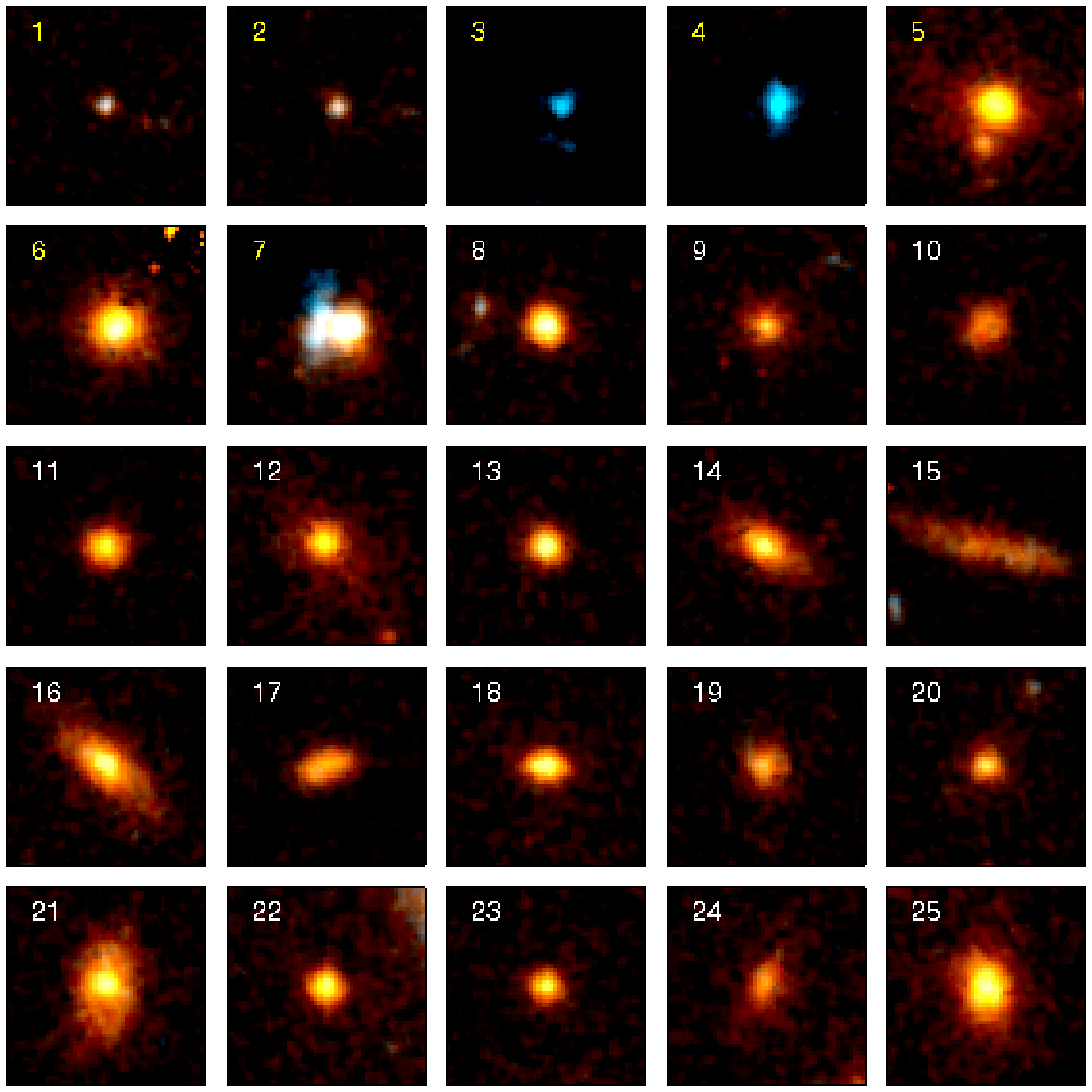}}
\centering
\caption{(Left panel): Pseudo-color {\it HST} image of IDCS J1433.2+3306, 2' on a side (ACS/F814W and WFC3/F160W). The numbers correspond to the image cutouts in the right panel.  (Right panel): Pseudo-color {\it HST} image cutouts, 3.25'' on a side.  The first seven panels are the confirmed spectroscopic members, marked in yellow, while the remaining 18 are candidate members of the red sequence as discussed in \S5 and Figures 6 and 7.  Spectroscopic members \#3 and \#4 lie outside of the WFC3/F160W imaging and hence show only F814W images.  Spectroscopic member \#7 is X-ray detected and appears to be in a merger system.}\label{fig:f2}
\end{figure*}

\begin{figure*}[htp] 
\setlength\fboxsep{0.5pt}
\setlength\fboxrule{0.5pt}
\fbox{\includegraphics[width=.45\textwidth]{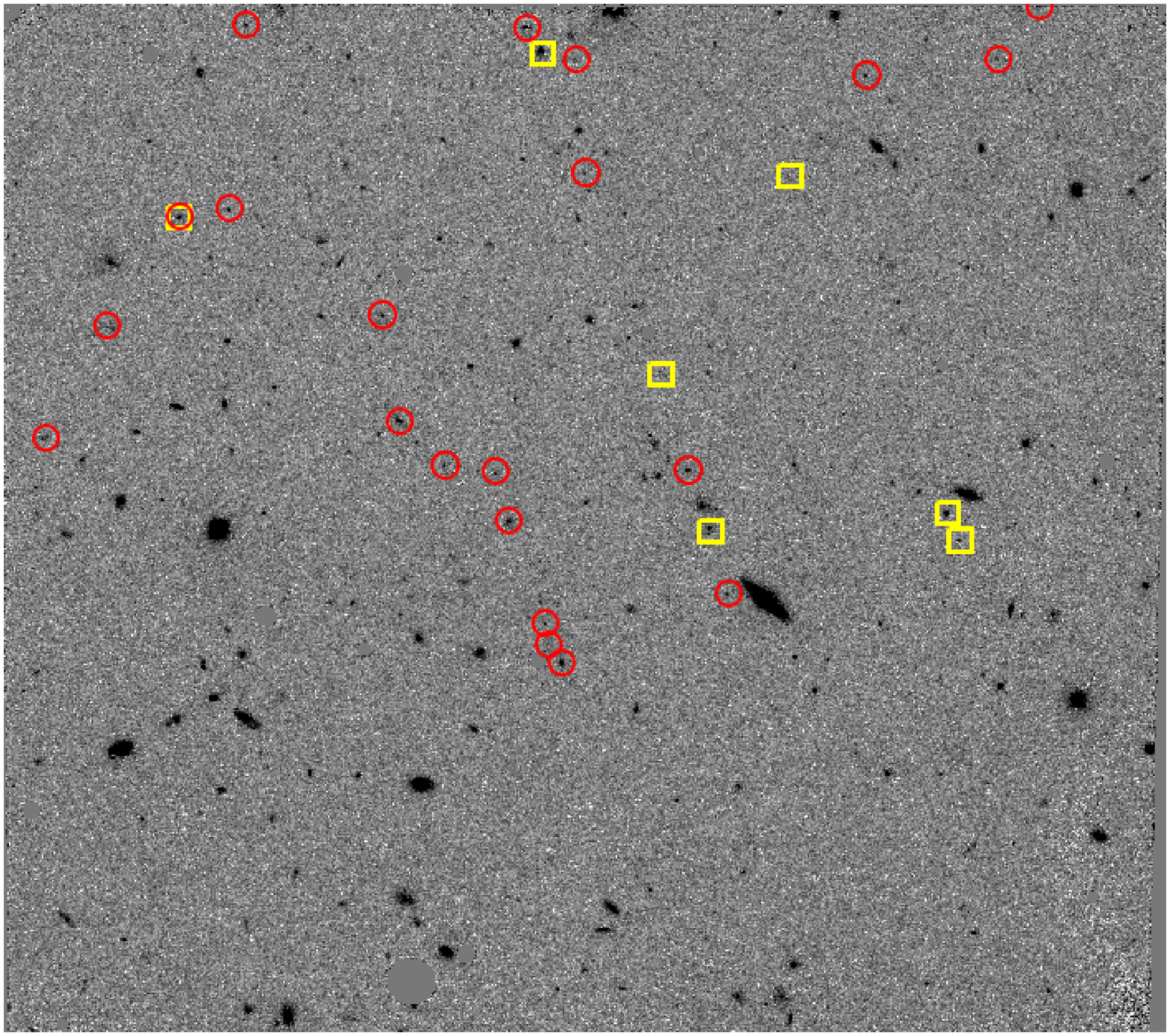}}
\fbox{\includegraphics[width=.45\textwidth]{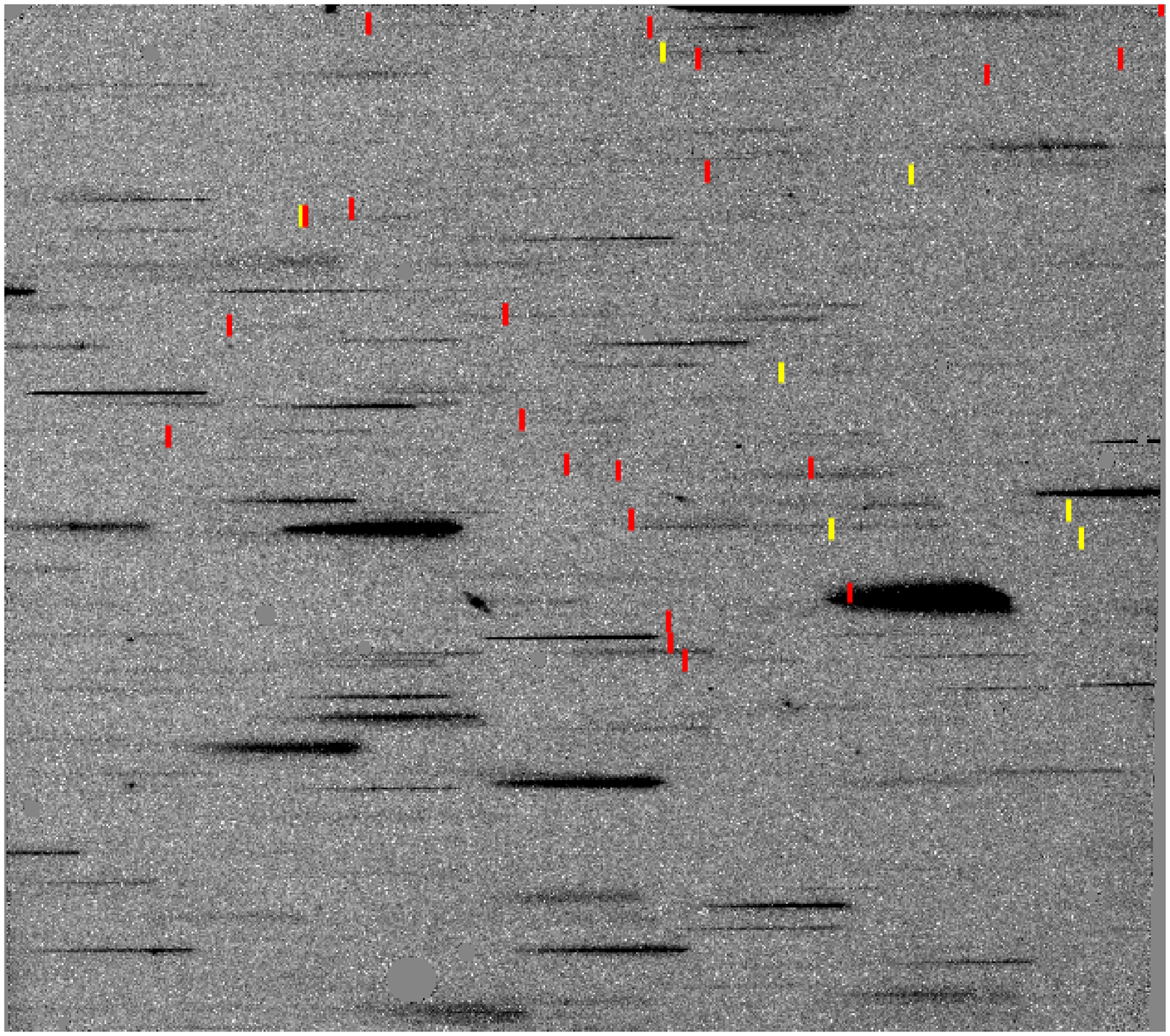}}
\fbox{\includegraphics[width=.45\textwidth]{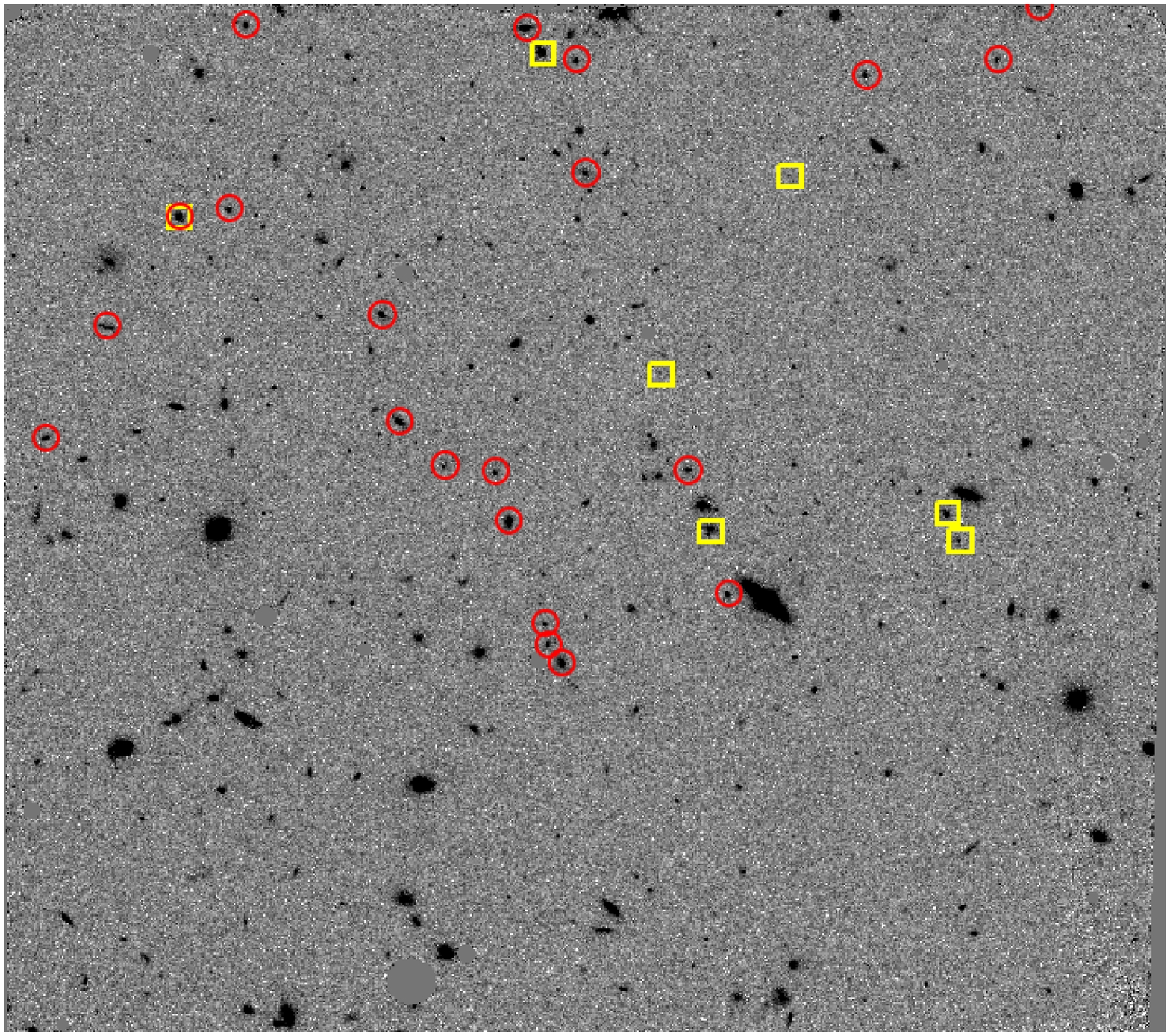}}
\fbox{\includegraphics[width=.45\textwidth]{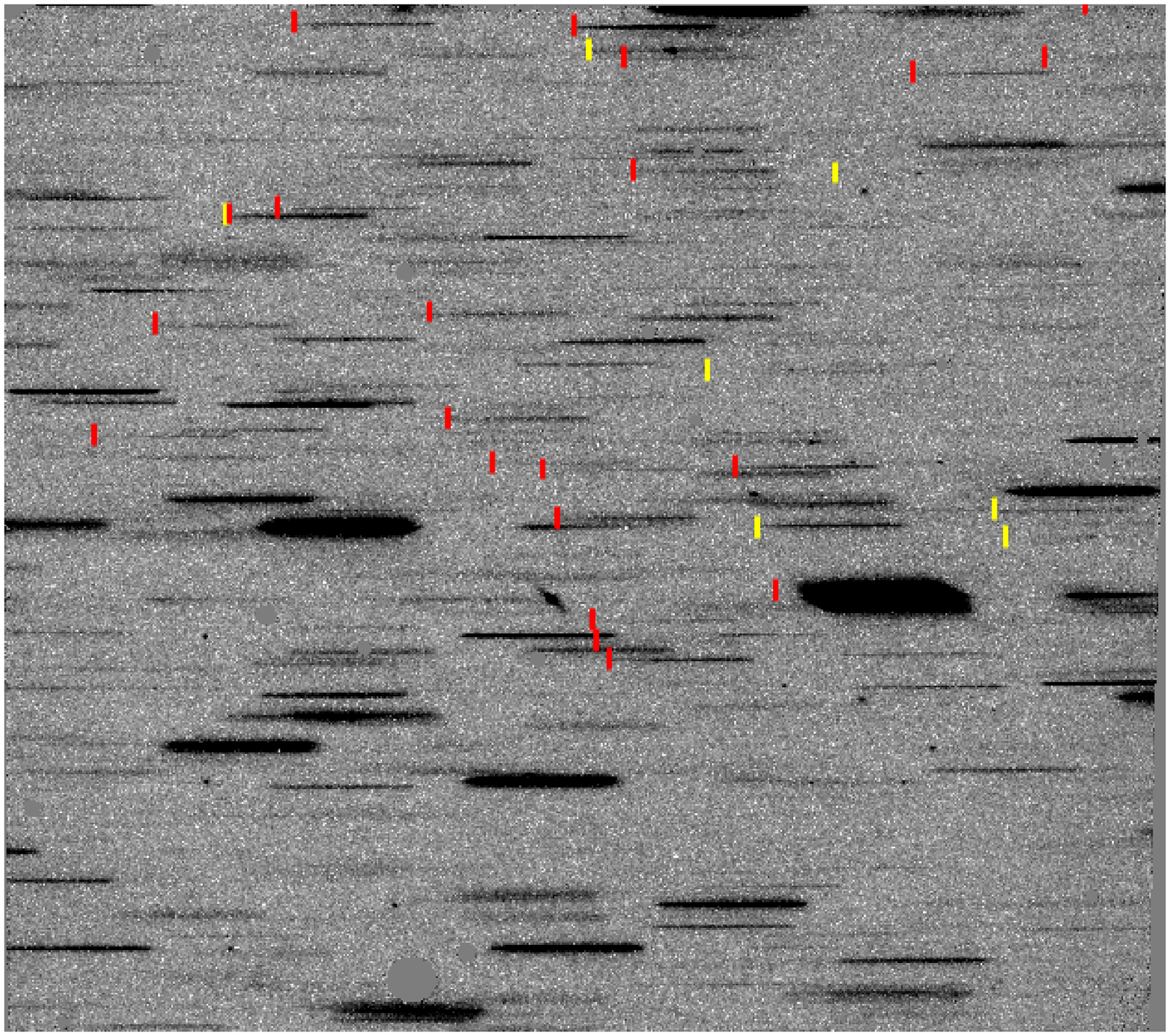}}
\centering
\caption{(Top left panel): F105W image from WFC3 of the cluster IDCS J1433.2+3306 with a field of view of 136$\arcsec \times 123\arcsec$.  This is a stacked image of 4x100s exposures, reduced with standard procedures.  Yellow boxes are spectroscopically confirmed members while red circles are candidate red sequence galaxies (see text and Figures 6 \& 7 for details).  (Top right panel): G102 2-D grism image from WFC3.  The image was reduced and stacked using a combination of aXe and Multidrizzle with a total exposure time of 11247 seconds.  The yellow and red tick marks signify the origin of the spectral trace for the spectroscopic members and red sequence candidates, respectively. (Bottom left panel): F140W image.  This is a stacked image of 4x100s exposures, reduced with standard procedures.  Markings are the same as in the top left panel.  (Bottom right panel): G141 2-D grism image with a total exposure time of 2011 seconds.  Markings are the same as in the top right panel.}\label{fig:f3}
\end{figure*}

\begin{figure*}[htp]
\includegraphics[]{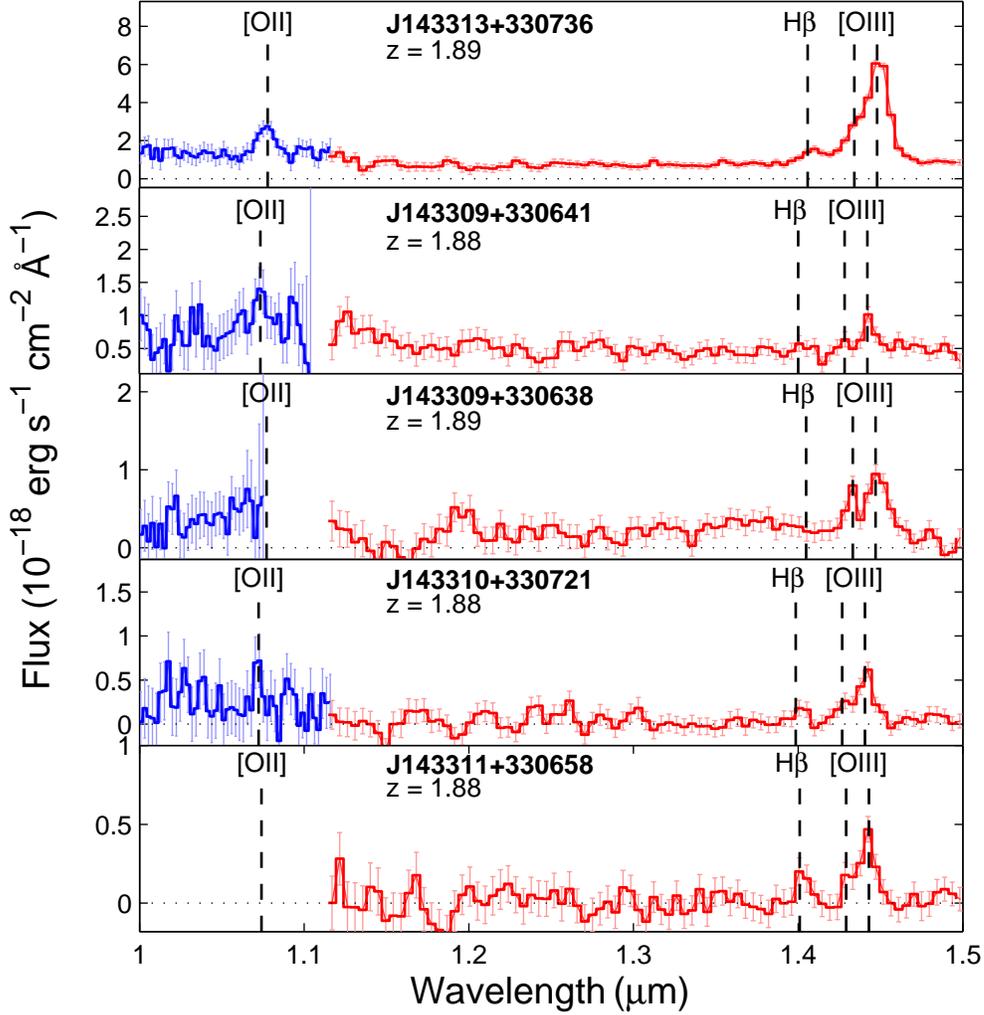}
\centering
\caption{WFC3 grism spectra of the five cluster members that exhibit emission lines.  The blue and red histograms are the spectra from the G102 and G141 grisms, respectively.  Emission features ([OII], H$\beta$, [OIII]4959, and [OIII]5007) are marked with black dashed vertical lines.  The top panel, J143313.1+330736, is an X-ray point source found in XBo\"{o}tes and a mid-IR selected active galaxy from Stern et al. (2005).}\label{fig:f4}
\end{figure*}

\begin{figure*}[htp]
\includegraphics[]{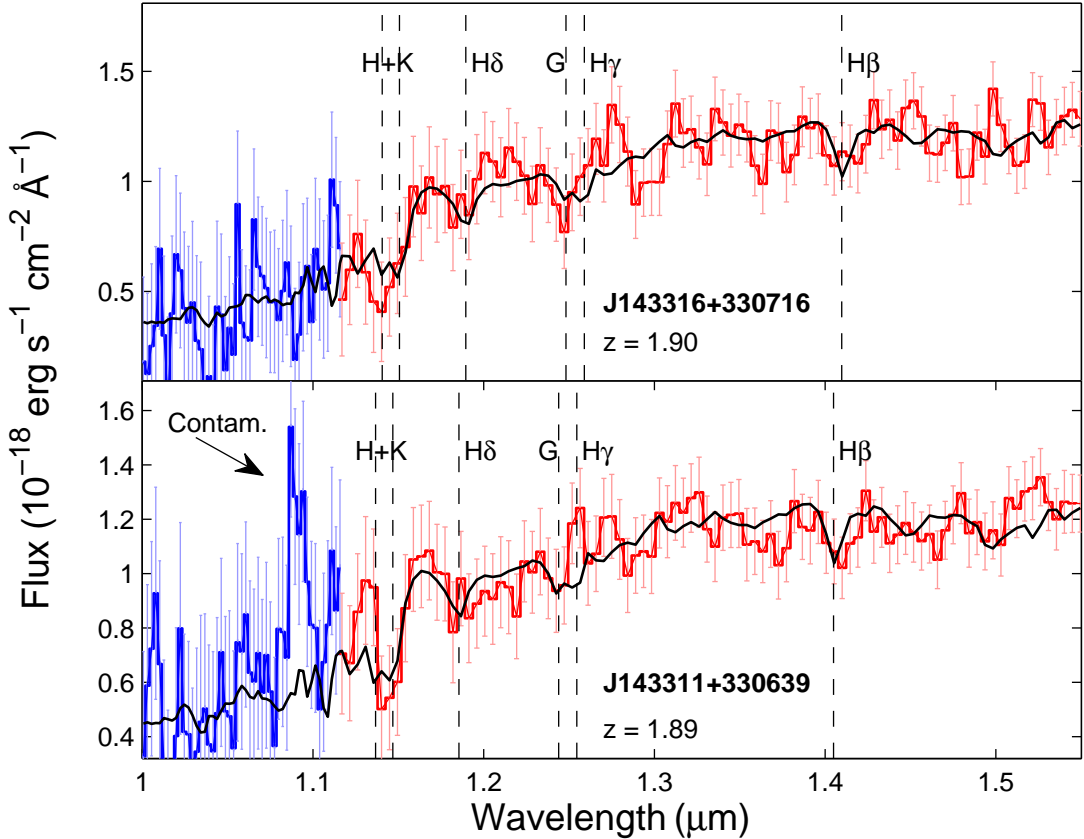}
\centering
\caption{WFC3 grism spectra of the two cluster members with early-type spectra.  The blue and red histograms are the spectra from the G102 and G141 grisms, respectively.  Absorption features (Ca H+K, H$\delta$, G-band, H$\gamma$, and H$\beta$) are marked with black dashed vertical lines.  Spectral collisions are quite common in slitless spectroscopy.  An arrow in the bottom panel marks a zeroth order contamination due to an overlapping spectrum which was masked in the spectral analysis.  The black solid line is the best fit model from the 2007 version of \citet{bruzual2003}, described in the text.  The best-fit model to the spectra in the top panel has an age of 0.9 Gyr, extinction A$_V$ = 0.8, and super-solar metallicity ($Z$ = 2.5$Z_{\odot}$).  The best-fit model to the spectra in the bottom panel has an age of 1.4 Gyr, extinction A$_V$ = 0.3, and solar metallicity.}\label{fig:f5}
\end{figure*}

\begin{figure*}[htp]
\includegraphics[width=.95\textwidth]{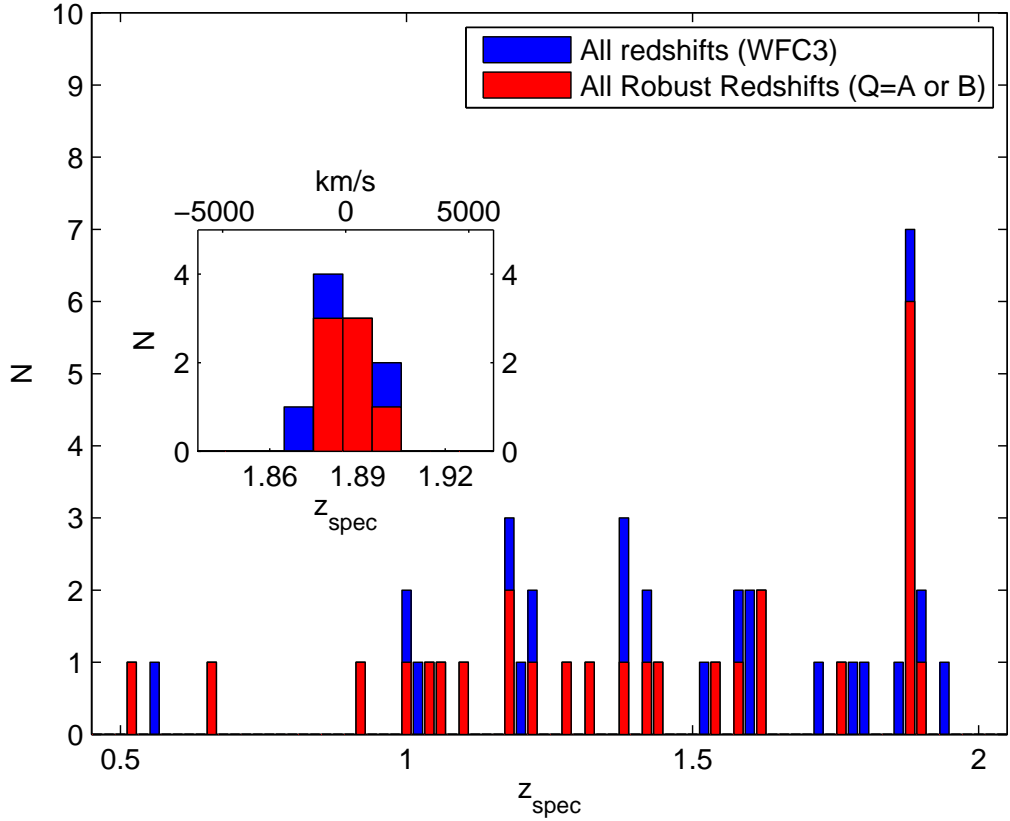}
\centering
\caption{Redshift histogram for the WFC3 spectroscopic observations.  The blue bars show all spectroscopic redshifts for all spectra while the red histogram shows only the robust redshifts (i.e., quality A or B).  The inset shows a detail of the redshift histogram near the cluster redshift.  The typical redshift error for a WFC3 grism measurement is ${\sigma}_{z}$ $\approx$ 0.01.}\label{fig:f6}
\end{figure*}

\begin{figure*}[htp]
\includegraphics[width=.95\textwidth]{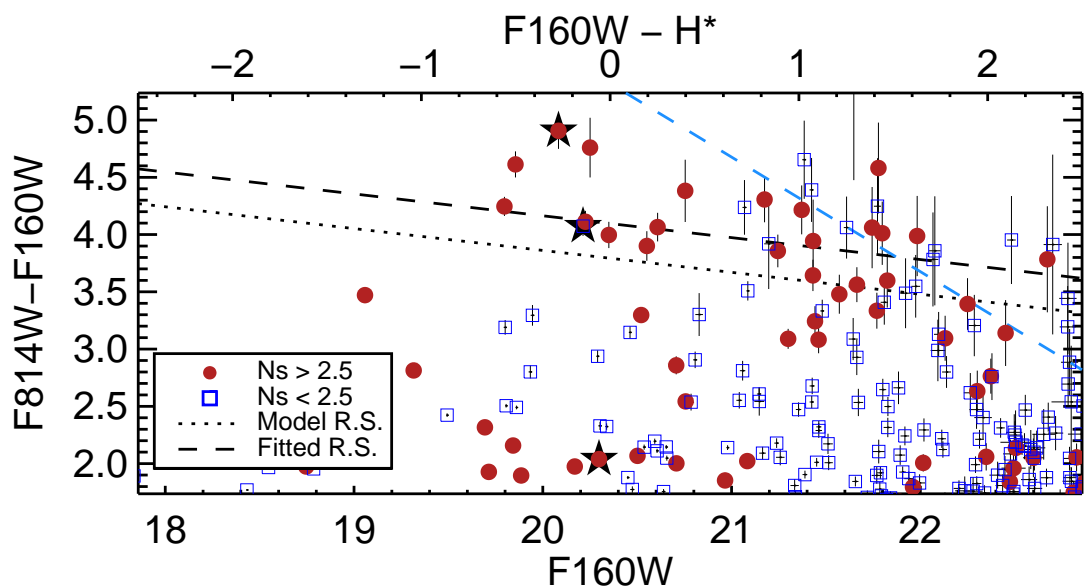}
\centering
\caption{Color-magnitude diagram of IDCS J1433.2+3306, created from ACS and WFC3 imaging. The red circles represent early-type galaxies ($n$ $>$ 2.5) while the blue squares denote late-type galaxies ($n$ $\le$ 2.5), as defined by the Sersi\c{c} index, $n$. The spectroscopically confirmed members are marked by the larger stars.  The black dotted line represents the expected color of a passively evolving red sequence of galaxies formed at $z$$_f$ = 3, with the slope based on observed Coma colors (Eisenhardt et al. 2007). The diagonal black dashed line is the fit to the red sequence galaxies. The diagonal dashed blue line represents the 5 $\sigma$ limit on the colors.  The bluest cluster member seen in this plot is an X-ray/mid-IR AGN.}\label{fig:f7}
\end{figure*}

\begin{figure*}[htp]
\includegraphics[width=.95\textwidth]{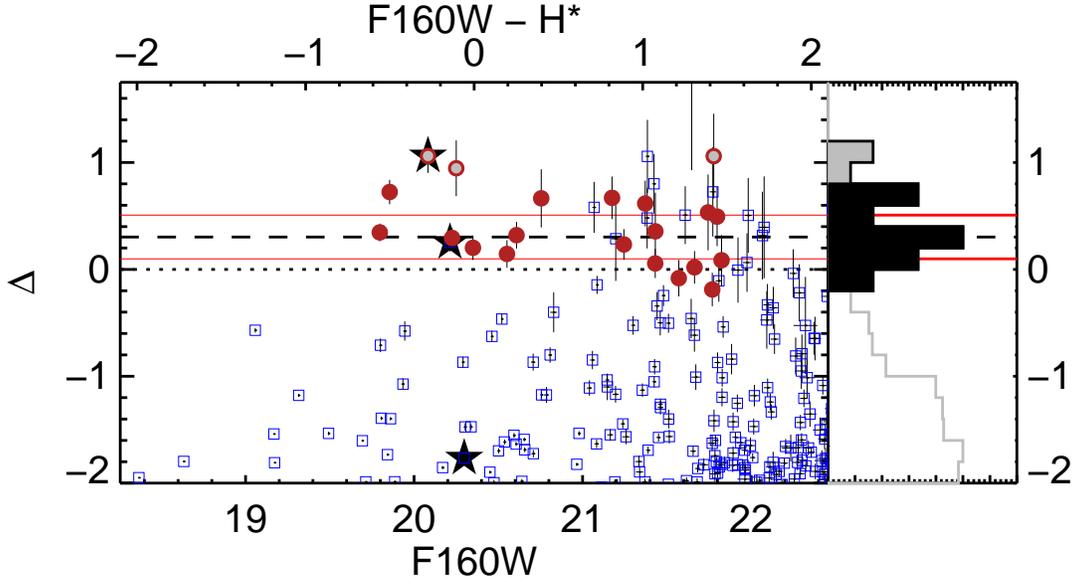}
\centering
\caption{The left panel of the plot is similar to the color-magnitude diagram in Figure 7.  The F814W - F160W colors have been zero pointed to the color predicted by a passively evolving red sequence of galaxies formed at $z$$_f$ = 3 with the slope base on observed Coma colors (Eisenhardt et al. 2007).  The red circles represent early-type galaxies ($n$ $>$ 2.5) which were selected by the magnitude and color cuts described in the text.  The grey circles were removed from the red sequence selection using a two median absolute deviation cut.  The solid red horizontal lines show the one ${\sigma}_{\rm int}$ intrinsic scatter above and below the central red sequence, and the black dashed line is the fit to the colors of the red sequence candidate galaxies.  The spectroscopically confirmed members are marked by the larger stars; only three of the members are red enough to appear in this plot.  One of the members was redder than the 2 $\sigma$ cut used to measure the red sequence while the member on the red sequence has a late-type morphology as measure by the Sersi\c{c} index, $n$ $\le$ 2.5.  The right side of the plot is a histogram stacked in color.}\label{fig:f8}
\end{figure*}

\clearpage
\begin{deluxetable}{cllllllcccc}
\rotate
\tabletypesize{\scriptsize}
\tablecaption{Spectroscopic Cluster Members}
\tablewidth{0pt}
\tablehead{
\colhead{Number\footnotemark[1]} & \colhead{ID} & \colhead{RA} & \colhead{Dec.} &  \colhead{$z$}  & \colhead{$\Delta$$z$} & \colhead{Q} & \colhead{F160W} & \colhead{F814W-F160W\footnotemark[5]}
 & \colhead{SFR$_{\rm H\beta}$} & \colhead{SFR$_{\rm [O\textsc{ii}]}$}\\
 \colhead{---} & \colhead{---} & \colhead{(J2000)} & \colhead{(J2000)} &  \colhead{---} & \colhead{---} & \colhead{---} & \colhead{(mag)} & \colhead{(mag)} &\colhead{(M$_{\odot}$ yr$^{-1}$)} & \colhead{(M$_{\odot}$ yr$^{-1}$)}
 }
\startdata
 1 & J143311.9+330658 & 14:33:11.98 & 33:06:58.0 &  1.88 & 0.01 & B  & 23.50 & 1.93 & 30 & ---\\
 2 & J143310.7+330721 & 14:33:10.75 & 33:07:21.6 & 1.88 & 0.01 & A  & 23.25 & 1.84 & 36 & 19\\
 3 & J143309.1+330638\footnotemark[3] & 14:33:09.14 & 33:06:38.0 & 1.89 & 0.01 & B  & --- & --- & $<$10\footnotemark[4] & ---\\
 4 & J143309.2+330641\footnotemark[3] & 14:33:09.26 & 33:06:41.2 & 1.88 & 0.01 & B  & --- & --- & 19 & 32\\
 5 & J143311.5+330639 & 14:33:11.51 & 33:06:39.4 & 1.89 & 0.01 & B  & 20.09 & 4.90 & --- & ---\\
 6 & J143316.5+330716 & 14:33:16.55 & 33:07:16.6 & 1.90 & 0.01 & B  & 20.22 & 4.07 & --- & ---\\
 7 & J143313.1+330736\footnotemark[2] & 14:33:13.11 & 33:07:36.4 & 1.89 & 0.01 & A  & 20.31 & 2.03 & AGN & AGN\\
 \hline
 --- & J143312.6+330642 & 14:33:12:68 & 33:06:42.4 & 1.87 & 0.01 & C & 21.43 & 2.68 & --- & ---\\
 --- & J143309.8+330711 & 14:33:09.86 & 33:07:11.7 & 1.90 & 0.01 & C & 22.47 & 1.61 & --- & ---\\
 --- & J143309.9+330628\footnotemark[3] & 14:33:09.90 & 33:06:28.1 & 1.88 & 0.01 & C & --- & --- & --- & ---\\

\enddata
\tablenotetext{1}{Number corresponding to Figure 2.}
\tablenotetext{2}{X-ray point source found in XBo\"{o}tes and mid-IR selected active galaxy from Stern et al. (2005).}
\tablenotetext{3}{Although detected in the F140W imaging obtained in support of the G141 grism observation, the source lies outside the WFC3 deeper imaging.}
\tablenotetext{4}{H$\beta$ emission was not detected.  Upper limit is the 68$^{th}$ percentile of the measured star formation rate for 1000 realizations of the data assuming Gaussian errors.}
\tablenotetext{5}{At $z = 1.89$, Mg\textsc{ii} falls within the F814W filter and [O\textsc{iii}] within the F160W filter.}

\end{deluxetable}

\begin{deluxetable}{cllc}
\tabletypesize{\tiny}
\tablecaption{Photometric Candidate Cluster Members}
\tablewidth{0pt}
\tablehead{
 \colhead{ID} & \colhead{RA} & \colhead{Dec.} &  \colhead{Photo-$z$} \\
\colhead{---} & \colhead{(J2000)} & \colhead{(J2000)} &  \colhead{---}
 }
\startdata
J143316.5+330716 & 14:33:16.54 & 33:07:16.5 & 1.85\footnotemark[1]\\
J143314.6+330704 & 14:33:14.61 & 33:07:04.7 & 1.94\\
J143305.5+330657 & 14:33:05.52 & 33:06:57.5 & 1.91\\
J143315.0+330722 & 14:33:14.96 & 33:07:22.7 & 1.80\\
J143304.6+330726 & 14:33:04.56 & 33:07:26.5 & 1.78\\
J143308.6+330747 & 14:33:08.65 & 33:07:47.7 & 1.89\\
J143313.5+330645 & 14:33:13.52 & 33:06:45.9 & 1.81\\
J143311.9+330536 & 14:33:11.90 & 33:05:36.6 & 1.79\\
J143315.6+330723 & 14:33:15.61 & 33:07:23.8 & 1.74\\
J143313.1+330634 & 14:33:13.08 & 33:06:34.1 & 1.90\\
J143305.4+330609 & 14:33:05.37 & 33:06:09.6 & 1.77\\
J143312.0+330710 & 14:33:12.01 & 33:07:10.2 & 1.80\\
J143308.8+330735 & 14:33:08.78 & 33:07:35.4 & 1.92\\
J143309.8+330645 & 14:33:09.77 & 33:06:45.1 & 1.81\\
J143312.7+330722 & 14:33:12.73 & 33:07:22.0 & 1.91\\
J143312.1+330649 & 14:33:12.06 & 33:06:49.4 & 1.88\\
J143316.0+330717 & 14:33:16.03 & 33:07:17.3 & 1.83\\
J143313.0+330504 & 14:33:12.98 & 33:05:04.1 & 1.78\\
J143312.9+330624 & 14:33:12.94 & 33:06:24.1 & 1.92\\
J143307.7+330823 & 14:33:07.69 & 33:08:23.1 & 1.73\\
J143310.6+330641 & 14:33:10.57 & 33:06:41.5 & 1.83\\
J143311.5+330641 & 14:33:11.54 & 33:06:41.0 & 1.88\footnotemark[2]\\
J143304.8+330654 & 14:33:04.78 & 33:06:54.4 & 1.76\\
J143305.0+330730 & 14:33:04.98 & 33:07:30.6 & 1.74\\
J143320.7+330732 & 14:33:20.69 & 33:07:32.1 & 1.75\\
J143317.8+330650 & 14:33:17.83 & 33:06:50.2 & 1.95\\
J143317.9+330641 & 14:33:17.90 & 33:06:41.1 & 1.70\\
J143316.4+330557 & 14:33:16.36 & 33:05:58.0 & 1.76\\
J143316.9+330548 & 14:33:16.87 & 33:05:48.1 & 1.71\\
J143310.0+330733 & 14:33:10.01 & 33:07:33.3 & 1.81\\
J143310.7+330456 & 14:33:10.68 & 33:04:57.0 & 1.68\\
J143309.9+330822 & 14:33:09.87 & 33:08:23.0 & 1.71\\
J143314.7+330700 & 14:33:14.73 & 33:07:00.8 & 1.69\\
J143311.7+330646 & 14:33:11.72 & 33:06:46.4 & 1.66\\
J143306.8+330629 & 14:33:06.75 & 33:06:29.0 & 1.64\\
J143313.2+330738 & 14:33:13.20 & 33:07:38.2 & 1.85\footnotemark[3]\\
J143308.8+330508 & 14:33:08.76 & 33:05:08.3 & 1.62\\
J143309.1+330553 & 14:33:09.08 & 33:05:53.6 & 1.64\\
J143310.5+330620 & 14:33:10.50 & 33:06:20.0 & 1.71\\
J143315.9+330739 & 14:33:15.93 & 33:07:39.8 & 1.61\\
J143317.7+330816 & 14:33:17.66 & 33:08:16.6 & 2.00\\
J143317.5+330647 & 14:33:17.47 & 33:06:47.7 & 1.62\\
J143309.7+330703 & 14:33:09.68 & 33:07:03.4 & 1.95\\

\enddata

\tablenotetext{1}{Match with spec-z member J143311.9+330658 ($<$ 3$\arcsec$)}
\tablenotetext{2}{Match with spec-z member J143311.5+330639 ($<$ 3$\arcsec$)}
\tablenotetext{3}{Match with spec-z member J143313.1+330736 ($<$ 3$\arcsec$)}

\end{deluxetable}

\begin{deluxetable}{cllccc}
\tabletypesize{\scriptsize}
\tablecaption{Red Sequence Candidate Cluster Members}
\tablewidth{0pt}
\tablehead{
 \colhead{ID} & \colhead{RA} & \colhead{Dec.} &  \colhead{F160W} & \colhead{F814W-F160W\footnotemark[1]} & \colhead{Photo-$z$} \\
\colhead{---} & \colhead{(J2000)} & \colhead{(J2000)} & \colhead{(mag)} & \colhead{(mag)} &  \colhead{---}
 }
\startdata
J143312.9+330623 & 14:33:12.91 & 33:06:23.4 & 20.23 & 4.10 & 1.92\\
J143313.0+330625 & 14:33:13.04 & 33:06:25.6 & 21.44 & 3.94 & ---\\
J143313.1+330628 & 14:33:13.07 & 33:06:28.1 & 21.80 & 4.00 & ---\\
J143311.3+330631 & 14:33:11.34 & 33:06:31.6 & 21.38 & 4.21 & ---\\
J143313.4+330640 & 14:33:13.42 & 33:06:40.3 & 20.36 & 3.99 & 2.07\\
J143313.5+330646 & 14:33:13.55 & 33:06:46.1 & 21.83 & 3.59 & 1.81\\
J143314.0+330646 & 14:33:14.03 & 33:06:46.9 & 21.82 & 3.40 & ---\\
J143311.7+330646 & 14:33:11.72 & 33:06:46.5 & 21.25 & 3.85 & 1.66\\
J143322.3+330650 & 14:33:22.30 & 33:06:50.2 & 21.67 & 3.56 & ---\\
J143325.7+330651 & 14:33:25.67 & 33:06:51.7 & 19.86 & 4.61 & ---\\
J143314.5+330652 & 14:33:14.45 & 33:06:52.2 & 20.56 & 3.89 & ---\\
J143320.5+330710 & 14:33:20.54 & 33:07:10.0 & 21.18 & 4.30 & ---\\
J143316.1+330717 & 14:33:16.08 & 33:07:17.5 & 21.44 & 3.64 & 1.83\\
J143312.7+330721 & 14:33:12.69 & 33:07:21.8 & 20.76 & 4.37 & 1.91\\
J143312.8+330735 & 14:33:12.78 & 33:07:35.4 & 21.75 & 4.05 & ---\\
J143313.3+330739 & 14:33:13.25 & 33:07:39.2 & 19.80 & 4.24 & 1.85\\
J143308.4+330741 & 14:33:08.38 & 33:07:41.8 & 20.61 & 4.06 & ---\\
J143315.5+330834 & 14:33:15.46 & 33:08:34.7 & 21.58 & 3.47 & ---\\

\enddata

\tablenotetext{1}{At $z = 1.89$, Mg\textsc{ii} falls within the F814W filter and [O\textsc{iii}] within the F160W filter.}

\end{deluxetable}

\end{document}